\documentclass[twocolumn,prb,showpacs]{revtex4}
\usepackage{graphicx}
\usepackage{bm}

\newcommand\ool{$H\!\!\parallel \!\! [001]$}
\newcommand\olo{$H\!\!\parallel \!\! [010]$}
\newcommand\loo{$H\!\!\parallel \!\! [100]$}
\newcommand\SEO{$\rm SrEr_2O_4$}
\newcommand\SLO{Sr$Ln_2$O$_4$}
\newcommand{\Kag}{Kagom\'{e}}
\newcommand\PRB[3]{Phys. Rev. B {\bf {#1}}, {#2} ({#3})}

\newcommand\PRL[3]{Phys. Rev. Lett. {\bf {#1}}, {#2} ({#3})}
\newcommand\JPCM[3]{J. Phys. Condens. Matter {\bf {#1}}, {#2} ({#3})}
\newcommand\JPSJ[3]{J. Phys. Soc. Japan {\bf {#1}}, {#2} ({#3})}

\newcommand\Tn{$T_{N}$}
\newcommand\afm{antiferromagnetic} 

\begin{document}
\title{Low-temperature magnetic ordering in SrEr$_2$O$_4$}
\author{O.~A.~Petrenko,$^1$  G.~Balakrishnan,$^1$ N.~R.~Wilson,$^1$ S.~de~Brion,$^2$ E.~Suard$^3$ and L.~C.~Chapon$^4$}
\affiliation{$^1$University of Warwick, Department of Physics, Coventry, CV4~7AL, UK}
\affiliation{$^2$Universit\'{e} Joseph Fourier and GHMFL-CNRS, BP 166, 38042 Grenoble cedex 9, France}
\affiliation{$^3$Institut Laue-Langevin, BP 156, 38042 Grenoble Cedex 9, France}
\affiliation{$^4$ISIS Facility, Rutherford Appleton Laboratory, Chilton, Didcot, OX11 0QX, UK}
\date{\today}
\begin{abstract}
\SEO\ has been characterised by low-temperature powder neutron diffraction, as well as single crystal specific heat and magnetisation measurements.
Magnetisation measurements show that the magnetic system is highly anisotropic at temperature above ordering.
A magnetic field of 280~kOe applied at $T=1.6$~K does not overcome the anisotropic magnetisation and fails to fully saturate the system.
Long range antiferromagnetic ordering develops below $T_N=0.75$~K, identified by magnetic Bragg reflections with propagation vector ${\bf k}=0$ and a lambda anomaly in the specific heat.
The magnetic structure consists of ferromagnetic chains running along the \textit{c}~axis, two adjacent chains being stacked antiferromagnetically.
The moments point along the \textit{c} direction, but only one of the two crystallographically in-equivalent Er sites has a sizeable ordered magnetic moment, 4.5~$\rm \mu_B$ at 0.55~K.
The magnetic properties of \SEO\ are discussed in terms of the interplay between the low-dimensionality, competing exchange interactions, dipolar interactions and low lying crystal field levels.
\end{abstract}
\maketitle

\section{Introduction}
Among the magnetic materials in which geometrical frustration plays an important role, the most intensely studied are the systems with triangular motif lattices such as stacked triangular lattices~\cite{Triangular}, two-dimensional (2D) and three-dimensional (3D) \Kag\ lattices~\cite{Kagome,GGG}, and 3D networks of corner-sharing (mostly found in pyrochlore~\cite{pyro} and spinel~\cite{spinel} compounds) or edge-sharing tetrahedra~\cite{FCC}.

The honeycomb lattice, where magnetically active ions form a 2D network of edge-sharing hexagons, is not a geometrically frustrated lattice if only the first neighbour exchange interactions are considered.
However, since it has the smallest possible 2D coordination number ($z=3$), quantum fluctuations are expected to be larger than for a square lattice, which justifies the considerable theoretical interest in the honeycomb lattice.  
Frustration can still originate from competing exchange interactions due to next-nearest neighbour interactions or additional coupling out of the plane.
The honeycomb lattice~\cite{note} can be viewed as composed of two interlacing triangular sublattices with each site having its three nearest neighbours on the other sublattice and its six second-nearest neighbours on its own sublattice~\cite{honey_theory}.

Recently, Karunadasa {\it et al.}~\cite{Karun} reported on the crystal structure and magnetic properties of the family of materials with general formula \SLO, where $Ln$ is Gd, Dy, Ho, Er, Tm and Yb.
These materials crystallise in the CaFe$_2$O$_4$ (calcium ferrite)-type structure~\cite{Decker} and have been studied for a number of years now.
What had not been realised prior to Karunadasa's publication, however, is the fact that in these compounds (and in several other isostructural compounds, such as Ba$Ln_2$O$_4$~\cite{Doi}, Pr$_2$BaO$_4$~\cite{Felner}) the magnetic $Ln$ ions are linked through a network of triangles and hexagons, where geometrical frustration can arise, provided that some exchange interactions are \afm.
In these systems, the honeycomb lattice is distorted in that the hexagons of magnetic rare-earth sites, octahedrally coordinated by oxygen ions, adopt a buckled configuration.
The honeycomb layers are stacked along the $c$~axis and each of the magnetic sites in one layer is connected to one of the sites in an adjacent honeycomb layer, forming a triangular network perpendicular to the honeycomb planes.
The magnetic susceptibility measurements reported by Karunadasa for \SEO\ (SEO) present no sharp features or pronounced changes in slope down to at least 1.8~K.
Yet the value of the Curie-Weiss constant $\theta_{CW}$  is -13.5 K, which suggests the presence of relatively strong antiferromagnetic interactions in this system~\cite{Karun}.
This disparity between the ordering temperature and the scale of the exchange interactions is typical for frustrated systems, although a significant reduction of the transition temperature may also arise from the low-dimensionality of the magnetic lattice.

In the present work we report on the low-temperature properties of SEO, investigated by powder neutron diffraction and single crystal specific heat and magnetisation measurements.
The system is found to order magnetically at $T_N=0.75$~K with a collinear magnetic structure.
However, only one of the two Er sites possesses a sizeable magnetic moment of 4.5~$\rm \mu_B$ at $T=0.55$~K.
The magnetic moments point along the $c$~axis and along this direction form ferromagnetic chains, which are arranged antiferromagnetically with respect to one of the neighbouring chains.
These ordered ``double" chains are separated by chains which contain almost no ordered magnetic moments (less than 0.5~$\rm \mu_B$ at $T=0.55$~K).
It is very likely that long-range and short-range magnetic order coexist at all temperatures below \Tn.
Magnetisation measurements show that the material is highly anisotropic above \Tn, with an anisotropy energy not totally overcome by magnetic fields as high as 280~kOe.

\begin{figure}[tb]
\includegraphics[width=0.9\columnwidth]{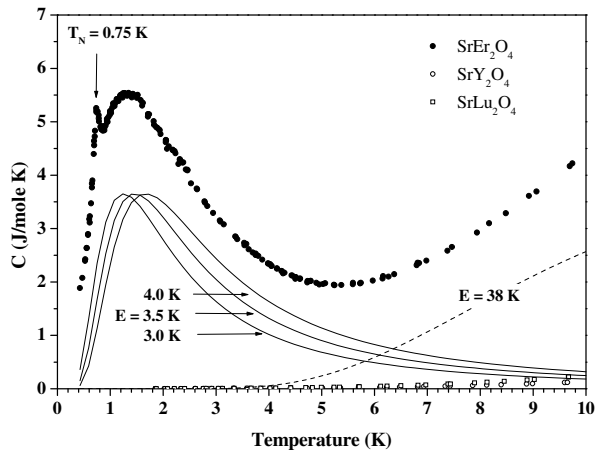}
\caption{\label{CvsT}
Temperature dependence of the specific heat of \SEO\ (solid symbols) and of the non-magnetic isostructural compounds SrLu$_2$O$_4$ and SrY$_2$O$_4$ (open symbols).
Solid and dashed lines are calculations of the specific heat Schottky anomaly expected for crystal field effects between a doublet ground state and a first excited doublet separated by energies indicated in the Figure (see text for details).}
\end{figure}

\section{Experimental details}
Polycrystalline samples of \SEO\ were prepared from the high-purity starting materials SrCO$_3$ and Er$_2$O$_3$ following Ref.~\onlinecite{Karun}.
X-ray diffraction measurements have confirmed the absence of any significant amount of impurities in the samples prepared.
High-resolution neutron powder diffraction experiments were conducted using a wavelength of 2.4~\AA\ on the D2B diffractometer at the Institut Laue-Langevin, France.
Diffraction patterns were collected in zero-field at various temperatures between 0.45 and 15~K using a $^3$He insert in a standard cryostat.
Rietveld refinements were performed with the FullProF program~\cite{FullP}.
Possible symmetry arrangements of the low-temperature magnetic structure were determined using representation analysis.

The single crystals of SEO were grown by the floating zone technique using an infra-red image furnace; the details of crystal growth of the entire \SLO\ family of compounds will be reported separately~\cite{Balakrishnan}.
The principal axes of the samples were determined using X-ray diffraction Laue photographs; for the magnetisation measurements the crystals were aligned to within an accuracy of better than 3$^{\circ}$.
Low temperature specific heat measurements were performed using a Quantum Design PPMS calorimeter equipped with a $^3$He option.
In order to estimate the lattice contribution to the specific heat we have measured the heat capacity of a single crystal of SrLu$_2$O$_4$ and SrY$_2$O$_4$, nonmagnetic compounds isostructural to SEO. 

Taking advantage of the high steady-state fields at the Grenoble High Magnetic Field Laboratory, we have taken magnetisation measurements for \loo, \olo\ and \ool\ at various temperatures above 1.6~K in applied fields of up to 280~kOe.
These measurements were performed using a standard extraction technique and a resistive magnet.
The same samples were used for more detailed investigations of the temperature dependence of magnetisation down to 1.4~K for different directions of an applied field using an Oxford Instruments vibrating sample magnetometer (VSM) in an applied field up to 120~kOe.

\section{Experimental results and Discussion}
\subsection{Heat capacity}
\begin{figure}[tb]
\includegraphics[width=0.9\columnwidth]{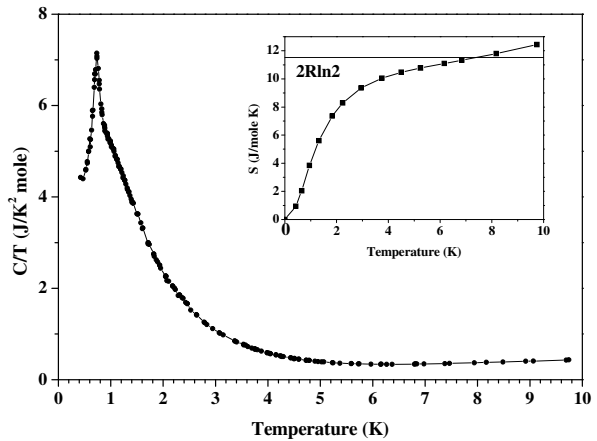} 
\caption{\label{CbyT}
Temperature dependence of the specific heat divided by temperature of \SEO.
The inset shows the temperature dependence of entropy, calculated as an area under the $C/T(T)$ curve, which has been extended linearly down to $T=0$~K.
The solid line indicates the position of $2R\ln{2}$, which corresponds to a magnetic contribution for a system with an effective $S=\frac{1}{2}$.}
\end{figure}

The temperature dependence of the specific heat of \SEO, shown in Figure~\ref{CvsT}, present two main features at low temperature.
First, a sharp lambda-type anomaly at about 0.75~K corresponding to a magnetic phase transition (as will be seen from the neutron diffraction results) and a much broader peak centred at around 1.3~K.
The broad peak resembles a feature that could be caused by a low-lying crystal-field (CF) excitation.
In order to test the latter possibility, we plot on the same graph the theoretical curves (solid lines) for a simple model, which contains a single excitation level separated from the ground state by 3~K, 3.5~K or 4~K.
The ground multiplet $^4$I$_{15/2}$ of Er$^{3+}$ is split into eight Kramers doublets in low symmetry crystals, such as \SEO.
We therefore have to presume that the degeneracy of both the ground state and the excited level is two.
As long as this is the case, a CF level cannot be solely responsible for the broad peak in specific heat -- the peak is about 50\% more intense than the doublets could generate regardless of their energy. 

\begin{figure}[tb]
\includegraphics[width=0.95\columnwidth]{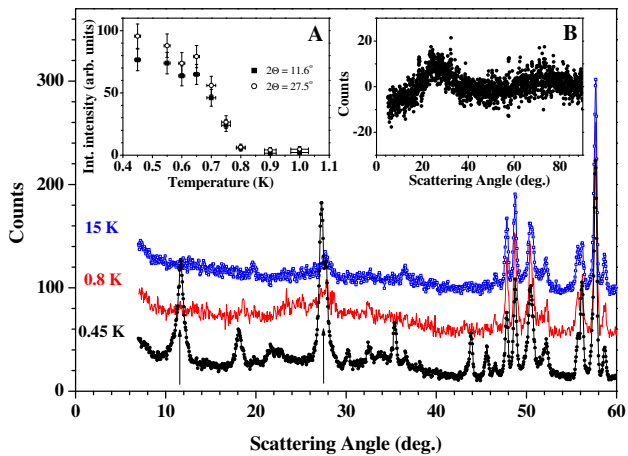}
\caption{\label{neutr1}(Colour online)
Low scattering angle fragment of the powder neutron diffraction data taken on \SEO\ using the D2B instrument at the ILL at 0.45~K, 0.8~K and 15~K.
The 0.8~K and 15~K data are shifted by 40 and 80 counts respectively for clarity.
The inset on the left shows the temperature dependence of the integrated intensity for the two magnetic peaks labelled with arrows on the main panel.
The inset on the right is a difference plot between the combined low temperature data ($T$ from 0.8~K to 1.5~K) and the $T=15$~K data.}
\end{figure}

Above 5~K the heat capacity begins to increase with temperature.
It is tempting to associate the observed rise with the lattice contribution (similar to what has been suggested for the Ba$Ln_2$O$_4$ compounds~\cite{Doi}).
However, the heat capacities of SrLu$_2$O$_4$ and SrY$_2$O$_4$, two nonmagnetic compounds isostructural with SEO, suggest that this is not the case.
In fact, the lattice contribution to the specific heat remains negligibly small even at $T=10$~K (see Fig.~\ref{CvsT}).
Unlike the lower temperature broad peak, the additional contribution to $C(T)$ at $T > 5$~K is probably caused by a CF level.
The dashed line in Figure~\ref{CvsT} was calculated for a CF level separated from the ground state by 38~K, which corresponds to the first peak observed at $\approx 3.3$~meV in our recent inelastic neutron scattering experiment~\cite{HET}.
A detailed analysis of the CF scheme in SEO goes beyond the scope of this paper.
It is still helpful, however, to consider the results of the specific heat measurements in terms of their implications for the magnetic properties of SEO.
For this purpose we plot in Figure~\ref{CbyT} the temperature dependence of the specific heat divided by temperature.
The area under the $C/T(T)$ curve represents the entropy, $S(T)$, which is shown in the inset of Figure~\ref{CbyT}.
At $T \approx 7$~K (above which temperature the input from the broad peak centred at 1.3~K is practically negligible) the entropy reaches the value of $2R\ln{2}$, corresponding to a purely magnetic contribution for an effective spin 1/2 system.
The entropy recovered by the ordering temperature  \Tn=0.75~K amounts to less than a quarter of the $2R\ln{2}$ value.

\subsection{Neutron diffraction}
From the neutron diffraction measurements in the paramagnetic phase, the space group of \SEO\ was confirmed as Pnam and the lattice parameters were found to be consistent with those previously reported \cite{Karun,Ukrainian}.
The refined values of $a$, $b$ and $c$ at 15 K were found to be 10.018(2)~\AA, 11.852(2)~\AA\ and 3.384(2)~\AA\ respectively.

Examples of the neutron diffraction patterns collected at various temperatures are shown in Figure~\ref{neutr1}.
Broad diffuse scattering peaks associated with the short-range magnetic order developing in SEO are observed at temperatures below 3~K (see inset B in Fig.~\ref{neutr1}).
This observation corroborates the suggestion that a broad peak at 1.3~K in the $C(T)$ curve is magnetic in origin.
The sharp peaks corresponding to long-range order appear in all the diffraction patterns taken below 0.8~K.
The intensity of these peaks has a pronounced temperature dependence (see inset A in Fig.~\ref{neutr1}), their positions, however, remained unchanged in the entire temperature range measured.
It was found that these peaks could be indexed with a propagation vector ${\bf k}=0$.

\begin{figure}[tb]
\begin{center}
\includegraphics[width=\columnwidth]{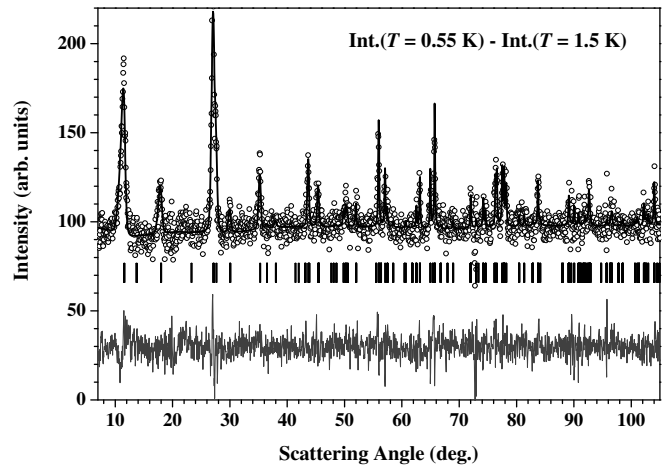}
\caption{\label{refinement}
Refined magnetic powder neutron diffraction data of \SEO\ taken as a difference between $T=0.55$~K and $T=1.5$~K.}
\end{center}
\end{figure}

The best fit to the data was found when the symmetry used was restricted to that of the basis vectors $\Gamma$7 for both Er ion sites.
At $T=0.55$~K the fit has a good agreement factor for the magnetic phase of $R_{mag}=15.5$\% and is shown in Figure~\ref{refinement}.
At the lowest experimentally available temperature of 0.45~K the agreement becomes less satisfactory, as the diffuse intensity observed at around 21 degrees and 34 degrees in scattering angle cannot be accounted for in a  ${\bf k}=0$ model.
In SEO, the presence of significant temperature dependant diffuse scattering with a pronounced structure factor allows a portion of the sample to support short range magnetic correlations, while the rest becomes long range ordered.

\begin{figure}[tb]
\includegraphics[width=\columnwidth]{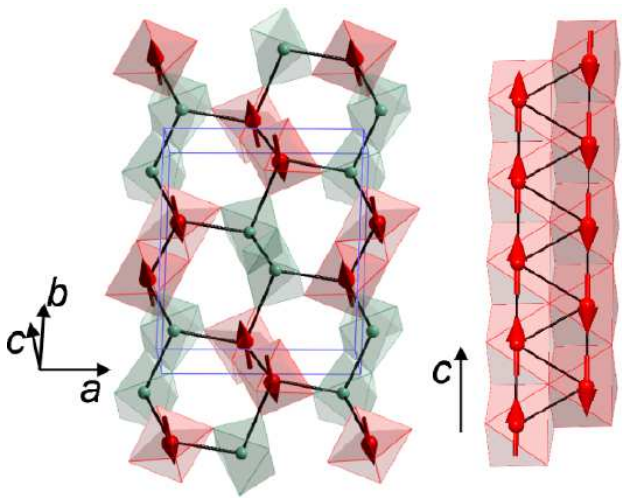}
\caption{\label{Structure} (Colour online)
Magnetic structure of \SEO\ as determined from Rietveld refinements of the neutron diffraction pattern at $T=0.55$~K.
The same structure is shown twice to emphasise different arrangement of the magnetic moments along the $c$~axis and with respect to the hexagons in the plane.
Two different Er sites and their surroundings are shown in different colour.
Only one of the sites carries a significant magnetic moment.}
\end{figure}

The proposed structure of the magnetic moments derived from the refinements and the observed magnetic extinction rules is shown in Figure~\ref{Structure}.
The two panels give the view parallel and perpendicular to the $c$~axis.
Since the $R_1$ and $R_2$ sites in the magnetic space group Pnam have mirror symmetry, the magnetic moments are either in the $ab$~plane or along the $c$~axis.
It should be concluded that the moments in SEO align along the $c$~axis, as no $(00l)$ reflections were observed.
They align ferromagnetically with their nearest neighbour, and adjacent chains align antiferromagnetically.
The Er$_1$ site has a magnetic moment of 4.5~$\rm \mu_B$, whereas the second Er$_2$ site has a much reduced moment of less than 0.5~$\rm \mu_B$.
Within our model the magnetic moments on Er$_1$ and Er$_2$ sites can be swapped without changing the calculated powder diffraction pattern.

Although the reasons for a large difference in the magnetic moment value on the two Er sites are likely to be related to the low lying CF levels, they cannot be solely responsible for the observed effects, as they cannot split further the Kramers doublet.
The involvement of exchange interactions and possibly dipolar forces must be presumed.
It is interesting to note that the proposed magnetic structure is favourable in terms of the dipolar interactions.
If only dipole-dipole interactions between the Er ions are considered (presuming equal moments on both sites) then a collinear model with the moments along the $c$~axis has one of the lowest possible energies.

\subsection{Magnetisation measurements}
\begin{figure}[tb]
\includegraphics[width=0.95\columnwidth]{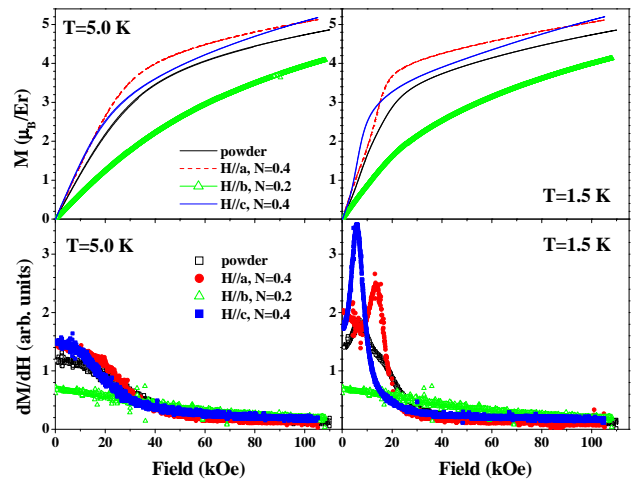}
\caption{\label{VSM}(Colour online)
Field dependence of magnetisation $M$ (top panels) and magnetic susceptibility $dM/dH$ (bottom panels) measured for \loo, \olo\ and \ool\ at 5~K (left panels) and 1.5~K (right panels).
$N$ denotes the demagnetisation factor used for calculation of the internal field.
The data for a powder sample are also shown for comparison.}
\end{figure}

The results of the VSM magnetisation measurements in lower-field are shown in Figure~\ref{VSM}.
The magnetisation is highly anisotropic at all measured temperatures and it does not reach the values predicted from the high-temperature susceptibility measurements of about 9~$\rm  \mu_B$ per Er ion~\cite{Karun}.
The magnetisation curves for different directions of the applied field are highly non-linear and have multiple crossing points (see Fig.~\ref{VSM}).
In this respect the definition of the easy-axis is field dependent: while the $b$~axis remains a hard direction in any field, the $c$~axis seems to be an easy direction in a weak field (up to 15~kOe), but becomes less favourable in higher fields up to about 90~kOe where the $M_{H\parallel [001]}(H)$ curve again crosses the $M_{H\parallel [100]}(H)$ curve.
Overall the magnetic anisotropy in SEO is best described as an easy-plane type, which is not uncommon for the Er$^{3+}$ ions.
It has to be borne in mind that all of this unusual behaviour is observed at 1.5~K, a temperature at which SEO remains magnetically disordered according to neutron powder diffraction data, which implies a strong CF influence.

At 1.5~K, the susceptibility curve dM/dH for a powder sample contains two broad maxima at 7~kOe and 17~kOe, while the single crystal curves for $H \! \parallel \! a$ and $H \! \parallel \! c$ only have a single, well pronounced maximum at 6~kOe and 14~kOe respectively.
It has been suggested~\cite{Karun} that a double peak behaviour of the dM/dH curve for a powder sample is indicative of the  presence of metamagnetism.
It is more likely, however, that a double peak feature in powder curves is simply a superposition of highly anisotropic curves for different directions of an applied magnetic field.
Usually broad maxima in susceptibility are the precursors of sharp peaks, which correspond to phase transitions at lower temperatures.
It would be very useful to extend the susceptibility measurements down to the magnetic ordering temperature in order to check if these maxima develop into field-induced phase transitions.

\begin{figure}[tb]
\includegraphics[width=0.95\columnwidth]{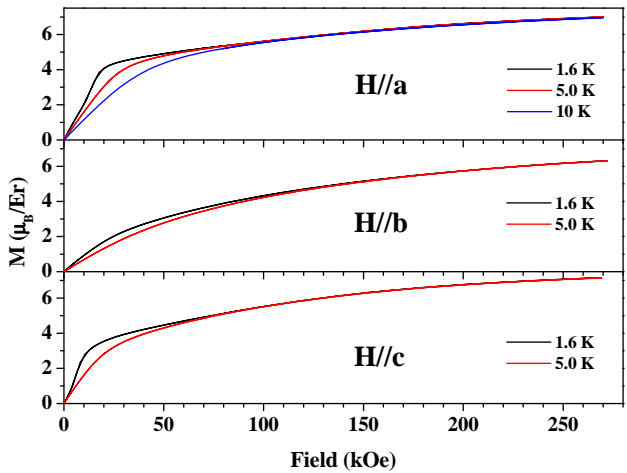}
\caption{\label{GHMFL}(Colour online)
Field dependence of magnetisation $M$ measured at the Grenoble High Magnetic Field Laboratory for \loo, \olo\ and \ool\ at different temperatures.
The data have been corrected for the demagnetisation effect.}
\end{figure}

In order to check if the full magnetic moment values expected for Er$^{3+}$ ion can be achieved in higher fields we have extended the magnetisation measurements up to 280~kOe.
The results shown in Figure~\ref{GHMFL} clearly illustrate that even a magnetic field of 280~kOe does not overcome the anisotropic magnetisation and fails to fully saturate the system.
The magnetisation, however, does not saturate in higher fields, it continues to increase with an applied field contrary to the conclusions made from the low-field measurements~\cite{Karun}.
A gradient of approximately 0.005~$\rm \mu_B / kOe$ and 0.008~$\rm \mu_B / kOe$ is observed in a field above 200~kOe for the $M(H)$ curves with \ool\ and \olo\ respectively.
The fact that average magnetic moment per Er ion in high fields is well above half of the value predicted from the high-temperature susceptibility measurements suggests that both of the two different Er sites carry a significant magnetic moment.

\section{Conclusions}
To summarise, neutron diffraction and specific heat measurements extended to lower temperatures prove that \SEO\ is a magnetically frustrated system which orders at 0.75 K~in zero field.
The moments are pointing along the $c$ axis, but the order is incomplete: only one of the two Er sites carries a sizeable magnetic moment.
Strong diffuse scattering is always present at sufficiently low temperature in zero field.
Further evidence of partial ordering and low-dimensionality is gathered from the heat capacity measurements, which show that three quarters of magnetic entropy are released only above the \Tn.
The behaviour above the ordering temperature is characterised by a significant degree of magnetic anisotropy of the easy-plane type, with the $b$~axis being a hard-magnetisation axis.
An application of a high magnetic field (280~kOe) fails to completely polarise the magnetic moments.

These results call for an introduction of a theoretical model which considers a competition between single ion anisotropy, dipole-dipole interactions and exchange interactions.
A combination of the three energies and of an external magnetic field is very likely to result in a complex $H-T$ phase diagram of \SEO\ at low temperatures.

The authors are indebted to M.~R.~Lees and M.~E.~Zhitomirsky for valuable discussions and A.~Richard for technical assistance.
We also acknowledge a collaboration with J.~Mercer and M.~L.~Plumer on a closely related project. 
This work is supported by the EPSRC research grant on Single Crystal Growth at Warwick.
The experimental time at the ILL was obtained through the EPSRC Super-D2B grant.
The work at Grenoble High Magnetic Field Laboratory is supported by the ``Transnational Access to Infrastructures - Specific Support Action" Program - Contract \. RITA-CT-2003-505474 of the European Commission.


\begin{thebibliography}{99}
\bibitem{Triangular} H.~Kawamura, \JPCM{10}{4707}{1998};
                                    M.~F.~Collins and O.~A.~Petrenko, Can. J. Phys. {\bf 75}, 605 (1997).
\bibitem{Kagome} A.~P.~Ramirez, G.~P.~Espinosa and A.~S.~Cooper, \PRL{64}{2070}{1990};
                                 C.~Broholm, G.~Aeppli, G.~P.~Espinosa, and A.~S.~Cooper, \PRL{65}{3173}{1990}.
\bibitem{GGG}  P.~Schiffer, A.~P.~Ramirez, D.~A.~Huse, P.~L.~Gammel, U.~Yaron, D.~J.~Bishop and  A.~J.~Valentino, \PRL{74}{2379}{1995};
                            P.~Schiffer, A.~P.~Ramirez, D.~A.~Huse and A.~J.~Valentino,  \PRL{73}{2500}{1994};
                            O.~A.~Petrenko, C.~Ritter, M.~Yethiraj  and D.~$\rm M^cK$~Paul, \PRL{80}{4570}{1998}.
\bibitem{pyro} for reviews on frustrated pyrochlore magnets see S.~T.~Bramwell and M.~J.~P.~Gingras, Science {\bf 294}, 1495 (2001);
                          J.~E.~Greedan, J. Mater. Chem. {\bf 11}, 37 (2001) and J. Alloys and Compounds, {\bf 408-412}, 444 (2006).
\bibitem{spinel} A.~P.~Ramirez, Handbook on Magnetic Materials, edited by K. J. H. Busch (Elsevier Science, Amsterdam, 2001), Vol. 13, p. 423.
\bibitem{FCC} K. Binder, \PRL{45}{811}{1980};
                           C.~L.~Henley, J. Applied Phys. {\bf 61}, 3962 (1987).
\bibitem{note} To avoid confusion, it should be stated that in some cases the term ``honeycomb structure" is used to describe the magnetic structure of Ising-like spins on a hexagonal lattice, where one third of the magnetic moments are aligned antiparallel to another third, while the residual third of the moments remain disordered~\cite{Ising}. In this article we use the term ``honeycomb structure" in a strictly crystallographic sense.
\bibitem{honey_theory} A.~Mattsson, P.~Frojdh and T.~Einarsson, \PRB{49}{3997}{1994}.
\bibitem{Karun} H.~Karunadasa, Q.~Huang, B.~G.~Ueland, J.~W.~Lynn, P.~Schiffer, K.~A.~Regan and R.J.~Cava, \PRB{71}{144414}{2005}.
\bibitem{Decker} B.~F.~Decker and J.~S.~Kasper, Acta Cryst. {\bf 10}, 332 (1957).
\bibitem{Doi} Y.~Doi,W.~Nakamori and Y.~Hinatsu, \JPCM{18}{333}{2006}.
\bibitem{Felner} I.~Felner, Y.~Yeshurun, G.~Hilscher, T.~Holubar, G.~Schaudy, U.~Yaron, O.~Cohen, Y.~Wolfus, E.~R.~Yacoby, L.~Klein, F.~H.~Potter, C.~S.~Rastomjee and R.~G.~Egdell, \PRB{46}{9132}{1992}.
\bibitem{FullP} J.~Rodriguez-Carvajal, Physica B, {\bf 192}, 55 (1993).
\bibitem{Balakrishnan} G.~Balakrishnan, T.~J.~Hayes, O.~A.~Petrenko and D.~$\rm M^cK$~Paul, in preparation (2008).
\bibitem{HET} T.~J.~Hayes, D.T.~Adroja and O.~A.~Petrenko, in preparation (2008).
\bibitem{Ukrainian}  L.~M.~Lopato and A.~E.~Kushchevskii, Ukrainskii Khimicheskii Zhurnal {\bf 39},  7 (1973).
\bibitem{Ising} M.~Mekata, \JPSJ{42}{76}{1977};  H.~Shiba, Prog. Theor. Phys. {\bf 64}, 466 (1980); F.~Matsubara and S.~Inawashiro, \JPSJ{56}{2666}{1987}.
\end{thebibliography}
\end{document}